\documentclass[11pt]{article}
\usepackage{moriond}

\def\be{\begin{equation}}
\def\ee{\end{equation}}
\def\bea{\begin{eqnarray}}
\def\eea{\end{eqnarray}}

\newcommand{\Photo}{\includegraphics[height=35mm]{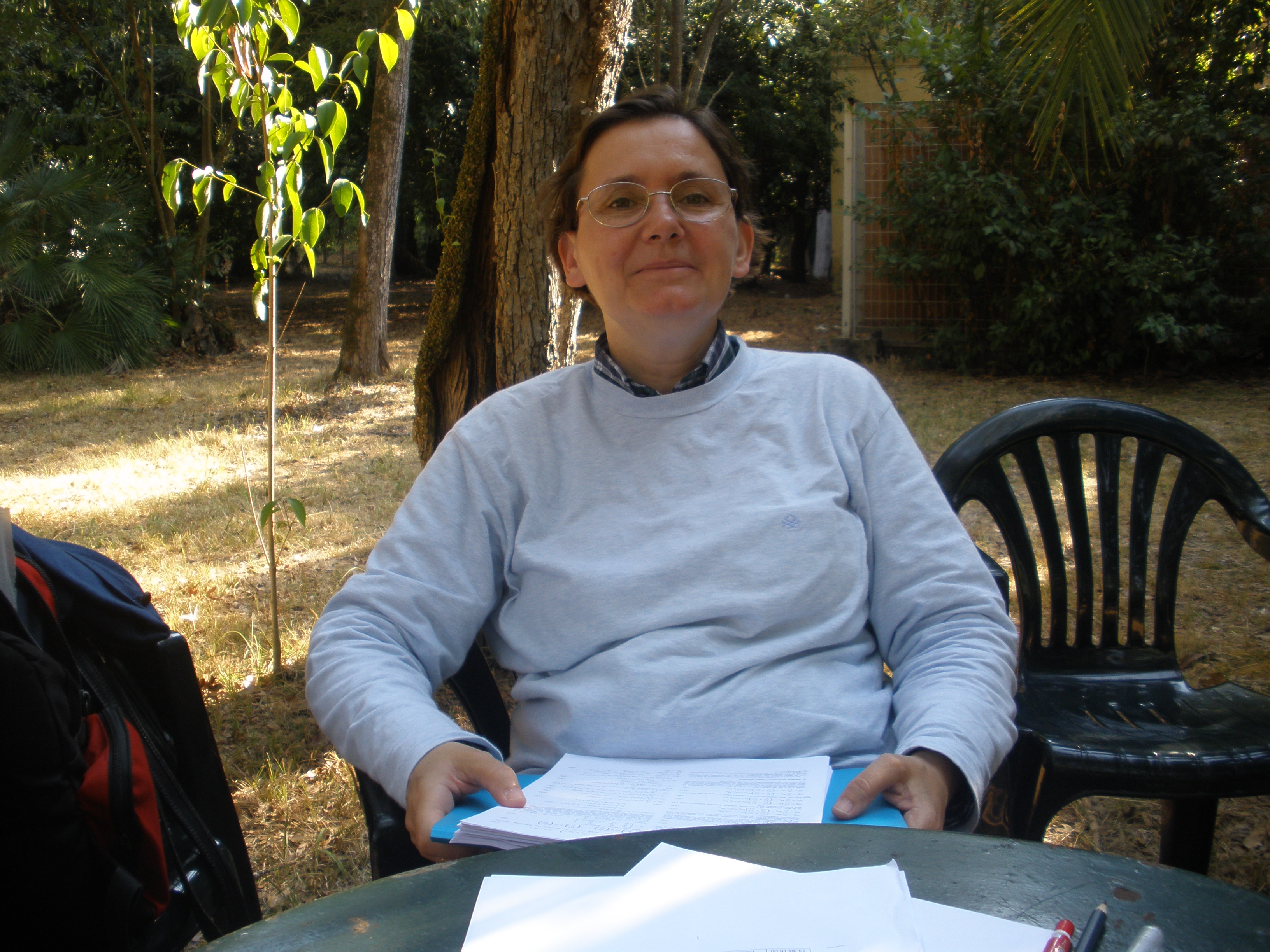}}

\begin{document}
\vspace*{4cm}
\title{INVARIANTS AND FLAVOUR IN THE GENERAL TWO-HIGGS DOUBLET MODEL}

\author{ M. N. REBELO 
}

\address{Universidade T\' ecnica de Lisboa, Centro de F\' \i sica Te\' orica
de Part\' \i culas (CFTP), Instituto Superior T\' ecnico, Av. Rovisco Pais, 
P-1049-001 Lisboa, Portugal.}

\maketitle\abstracts{We construct weak basis invariants which can give insight into
the physical implications of any flavour model, written in an arbitrary
weak basis (WB) in the context of 2HDM.
}

\section{Introduction}
The LHC has started the ``Higgs Era" with the discovery of a
scalar boson with a mass of approximately 126 GeV 
\cite{Aad:2012tfa}, \cite{Chatrchyan:2012ufa}, which seems to behave
like the Standard Model (SM) Higgs boson. There are very good motivations
to consider models with two Higgs doublets \cite{Branco:2011iw}, 
\cite{Gunion:1989we} despite the 
fact that the SM is a very successful theory. Most observations in the hadronic sector
are in agreement with the SM predictions apart from a few
anomalies and tensions. However, in the leptonic sector the SM has to
be extended in order to accommodate neutrino masses and leptonic mixing.
Furthermore, accounting for the baryon asymmetry of the Universe requires
new sources of CP violation thus providing a possible motivation to consider two
Higgs doublet models. Supersymmetry, if discovered, will also require the existence 
of two Higgs doublets. The discovery of a charged Higgs at the LHC in the future
would be an important step towards the experimental confirmation
of the need to extend the Higgs sector of the SM.

Models with two Higgs doublets have potentially large Higgs mediated
flavour changing neutral currents (FCNC). Experimentally FCNC are strongly
constrained. There are several possible ways of suppressing these currents.
Natural flavour conservation \cite{Glashow:1976nt} or the imposition of 
alignment \cite{Pich:2009sp} eliminate
tree level Higgs FCNC. In alternative, one may have FCNC at tree level suppressed
by small entries of the Cabibbo-- Kobayashi -- Maskawa matrix, $V_{CKM}$.
The first models of this type  based on a symmetry were built by Branco, Grimus
and Lavoura \cite{Branco:1996bq} and later on extended in Refs \cite{Botella:2009pq}, 
\cite{Botella:2011ne}. This talk is based on 
work done in collaboration with Botella and Branco \cite{Botella:2012ab}.

\section{General Framework}
The flavour structure of two Higgs doublet models is given by the Yukawa 
interactions:
\begin{equation}
L_Y = - \overline{Q^0_L} \ \Gamma_1 \Phi_1 d^0_R - \overline{Q^0_L}\  
\Gamma_2 \Phi_2 d^0_R - \overline{Q^0_L} \ \Delta_1 \tilde{ \Phi_1} u^0_R - 
\overline{Q^0_L} \ \Delta_2 \tilde{\Phi_2} u^0_R \ + \mbox{h. c.}
\label{1e2}
\end{equation} 
where $\Gamma_i$ and $\Delta_i$ denote the Yukawa couplings of the 
lefthanded quark doublets $Q^0_L$ to the righthanded quarks $d^0_R$,
$u^0_R$ and the Higgs doublets $\Phi_j$. The quark mass matrices generated
after spontaneous gauge symmetry breaking are given by:
\begin{eqnarray}
M_d = \frac{1}{\sqrt{2}} ( v_1  \Gamma_1 +
                           v_2  e^{i \alpha}   \Gamma_2 ), \quad 
M_u = \frac{1}{\sqrt{2}} ( v_1  \Delta_1 +
                           v_2  e^{-i \alpha} \Delta_2 ),
\label{mmmm}
\end{eqnarray}
where $v_i \equiv |<0|\phi^0_i|0>|$ and $\alpha$ denotes the relative phase 
of the vacuum expectation values (vevs) of the neutral components of 
$\Phi_i$. 
The matrices $M_d$,  $M_u$ are diagonalized by the usual bi-unitary transformations:
\begin{eqnarray}
U^\dagger_{dL} M_d U_{dR} = D_d \equiv {\mbox diag}\ (m_d, m_s, m_b) 
\label{umu}\\
U^\dagger_{uL} M_u U_{uR} = D_u \equiv {\mbox diag}\ (m_u, m_c, m_t)
\label{uct}
\end{eqnarray}
The neutral and the charged Higgs interactions obtained from Eq.~(\ref{1e2}) are of the form
\begin{eqnarray}
{\mathcal L}_Y (\mbox{quark, Higgs})& = & - \overline{d_L^0} \frac{1}{v}\,
[M_d H^0 + N_d^0 R + i N_d^0 I]\, d_R^0  - \nonumber \\
&-& \overline{{u}_{L}^{0}} \frac{1}{v}\, [M_u H^0 + N_u^0 R + i N_u^0 I] \,
u_R^{0}  - \label{rep}\\
& - &  \frac{\sqrt{2} H^+}{v} (\overline{{u}_{L}^{0}} N_d^0  \,  d_R^0 
- \overline{{u}_{R}^{0}} {N_u^0}^\dagger \,    d_L^0 ) + \mbox{h.c.} \nonumber 
\end{eqnarray}
where $v \equiv \sqrt{v_1^2 + v_2^2} \approx \mbox{246 GeV}$,  
 and $H^0$, $R$ are orthogonal combinations of the fields  $\rho_j$,
arising when one expands  \cite{Lee:1973iz}  the neutral scalar fields
around their vacuum expectation values, $ \phi^0_j =  \frac{e^{i \alpha_j}}{\sqrt{2}} 
(v_j + \rho_j + i \eta_j)$, choosing $H^0$ in such a way
that it has couplings to the quarks which are proportional
to the mass matrices, as can be seen from Eq.~(\ref{rep}). 
Similarly, $I$ denotes the linear combination
of $\eta_{j}$ orthogonal to the neutral Goldstone boson. 
The matrices $N_d^0$,  $N_u^0$  
are given by:
\begin{eqnarray}
N_d^0 = \frac{1}{\sqrt{2}} ( v_2  \Gamma_1 -
                           v_1 e^{i \alpha} \Gamma_2 ), \quad 
N_u^0 = \frac{1}{\sqrt{2}} ( v_2  \Delta_1 -
                           v_1 e^{-i \alpha} \Delta_2 )
\end{eqnarray}
The flavour structure of the quark sector of  two Higgs doublet models
is thus fully specified in terms of
the four matrices $M_d$,  $M_u$, $N_d^0$,  $N_u^0$.  
The physical neutral Higgs fields are combinations 
of  $H^0$, $R$ and $I$. Flavour changing neutral currents are controlled by $N^0_d$ and 
$N^0_u$. 

\section{Weak Basis Invariants}
The four flavour matrices $M_d$, $M_u$, $N_d^0$,  $N_u^0$
contain a large redundancy of parameters which results from 
the fact that under a weak basis (WB) transformation they change transforming as
\begin{eqnarray}
M_d \rightarrow M_d^\prime  = W_L^\dagger M_d W_R^d,  \qquad
M_u \rightarrow M_u^\prime  = W_L^\dagger M_u W_R^u, \nonumber \\
N_d^0  \rightarrow {N_d^0} ^\prime  = W_L^\dagger N_d^0  W_R^d,  \qquad
N_u^0 \rightarrow {N_u^0}^\prime  = W_L^\dagger N_u^0 W_R^u
\label{wbtra}
\end{eqnarray}
without altering their 
physical content. Different Lagrangians related to each other by WB transformations
describe the same physics. In view of the above redundancy, it is of great 
interest to construct WB invariants which can be very useful in the analysis of  
the physical content of the flavour sector of two Higgs doublet models 
\cite{Botella:2012ab} by following
the technique that was introduced in \cite{Bernabeu:1986fc} to the study  of CP violation in the
SM. This technique  was later generalized to many different scenarios, in particular to
the study of explicit CP violation in the scalar sector of multi-Higgs doublet models prior to 
gauge symmetry breaking \cite{Branco:2005em} as well as CP violation in the scalar sector 
after this breaking \cite{Lavoura:1994fv} and also taking into account both the scalar and the 
fermionic sector \cite{Botella:1994cs}.

In this framework, it is clear that one can build new WB basis invariants which do not 
arise in the SM by evaluating traces of blocks of matrices involving
the up and down quark sector, like for example $M_\gamma N_\gamma^{0 \dagger}$ or
$N_\gamma^0 N_\gamma^{0 \dagger}$. For definiteness let us consider the WB invariant 
tr($M_d N_d^{0 \dagger}$) and note that its physical significance becomes
transparent in the WB where $M_d$ is diagonal, real, since in this basis the
matrix $N_d^0$ already coincides with  the couplings to the  physical quarks. 
In this basis one has:
\begin{equation}
I_1 \equiv \mbox{tr} (M_d N_d^{0 \dagger}) = m_d(N_d^\ast)_{11} +
m_s(N_d^\ast)_{22} + m_b(N_b^\ast)_{33}
\label{nnn}
\end{equation}
We denote $N_d$, the matrix $N_d^0$ in the basis where it couples to the 
physical quarks. This invariant is not sensitive to  Higgs-mediated FCNC, but Im($I_1$)
is specially important, since it probes the phases of $(N_d)_{jj}$ which 
contribute to the electric dipole moment of down-type quarks. Obviously,
one can construct an analogous invariant for the up-quark sector, namely
tr($M_u N_u^{0 \dagger}$). Let us now consider a WB invariant
which is sensitive to the off-diagonal elements of $N_d$, namely:
 \begin{eqnarray}
I_2 \equiv \mbox{tr} \left[ M_d N_d^{0 \dagger}, M_d M_d^ \dagger \right]^2 
 = -2 m_d m_s (m_s^2 - m_d^2)^2 (N_d^\ast)_{12} (N_d^\ast)_{21} -
\nonumber \\
- 2 m_d m_b (m_b^2 - m_d^2)^2 (N_d^\ast)_{13} (N_d^\ast)_{31}  -
2 m_s m_b (m_b^2 - m_s^2)^2 (N_d^\ast)_{23} (N_d^\ast)_{32} ,
\end{eqnarray}
where we have kept the notation used in Eq.~(\ref{nnn}), having evaluated $I_2$
in the WB where $M_d$ is real and diagonal. WB invariants are also important
to study CP violation. In the SM a necessary and sufficient condition
for CP invariance is the vanishing of the WB invariant \cite{Bernabeu:1986fc} : 
\begin{eqnarray}
I_1^{CP} \equiv & \mbox{tr} \left[  H_u, H_d \right]^3  = 6i  (m^2_t - m^2_c)   
(m^2_t - m^2_u)   (m^2_c - m^2_u) \times \nonumber \\ 
& \times (m^2_b - m^2_s)   (m^2_b - m^2_d) (m^2_s - m^2_d)  \mbox{Im} Q_{uscb}
\label{eq31}
\end{eqnarray}
where $H_{d,u} \equiv (M_{d,u} M^\dagger _{d,u})$,   $Q$ stands for a rephasing invariant quartet of $V_{CKM}$,
defined by $Q_{\alpha i \beta j} \equiv V_{\alpha i}  V_{\beta j}  V^\ast_{\alpha j}  
V^\ast_{\beta i} $ ($\alpha \neq \beta$, $ i \neq j$) with
$V_{CKM} \equiv  U^\dagger_{uL}  U_{dL}$. The fact that  $V_{CKM}$ is not the identity
reflects the fact that 
$ U_{dL} \neq  U_{uL}$, i.e. that there is misalignment of the matrices
$H_d$, $H_u$ in flavour space.
For three generations $I_1^{CP} $ is proportional to det$\left[  H_u, H_d \right] $,
introduced in Ref.~\cite{Jarlskog:1985ht}.
In the present framework there are four matrices relevant for flavour rather than the
two mass matrices of the SM, therefore we can generalise  the
definition of $I_1^{CP} $ to diferent WB invariants sensitive to the
misalignment of different pairs of Hermitian matrices, such as, for example
$I_2^{CP} \equiv  \mbox{tr} \left[  H_u, H_{N_d^0} \right]^3 $ or
$I_3^{CP} \equiv  \mbox{tr} \left[  H_d,  H_{N_d^0}  \right]^3$,
with $  H_{N_d^0} \equiv N_d^0  N_d^{0 \dagger}$.

In the SM the lowest order WB invariant sensitive to
CP violation is given by Eq.~(\ref{eq31}) and has dimension twelve
in powers of mass. The richer flavour structure of models with two
Higgs doublets allows for lower order invariants sensitive to CP
violation, namely, for instance:
 \begin{equation}
I_9^{CP} \equiv \mbox{Im tr} \left[ M_d  N_d^{0 \dagger} M_d M^\dagger_d
M_u M^\dagger_u M_d M^\dagger_d \right] 
\end{equation}
In generic two Higgs doublet models one may even have lower order 
invariants sensitive to CP violation. However, with the imposition of a symmetry
in order to suppress Higgs mediated FCNC such lower other invariants
may become trivial.

Flavour symmetries and/or texture zeros reduce the number
of free parameters and have physical implications. However symmetries 
and textures are introduced
in a specific WB. Under a change of WB these will in principle cease to be apparent.
In this respect the computation of weak basis invariants is a very useful tool
to recognize properties related to special symmetries or textures that
may not be apparent. In Ref.~\cite{Botella:2012ab} the summary presented here
is extended and applied to two special cases: models of the type proposed
by Branco, Grimus and Lavoura  \cite{Branco:1996bq} and a special implementation of
nearest -- neighbour -- interaction (NNI) textures \cite{Fritzsch:1977vd}
in the context of two Higgs doublet models based on an Abelian 
symmetry\cite{Branco:2010tx}.

\section*{Acknowledgments} The author thanks the organizers of 
Rencontres de Moriond for the invitation to present this work anf for 
the very stimulating atmosphere.
The work presented in this Conference
was  partially supported by Funda\c c\~ ao 
para a Ci\^ encia e a Tecnologia (FCT, Portugal) through the projects
PTDC/FIS/098188/2008 and CFTP-FCT Unit 777 (PEst-OE/FIS/UI0777/2011)
which are partially funded through POCTI (FEDER),  by Marie Curie ITN "UNILHC" 
PITN-GA-2009-237920,
by  Accion Complementaria Luso-Espanhola
PORT2008--03, by European FEDER and FPA-2008-04002-E/PORTU by
Spanish  MICINN under grant FPA2008--02878 and GVPROMETEO 2010-056.
More recently we would like to aknowledge the support of CERN/FP/123580/2011.

\end{document}